\def\breakon{\end{multicols}\widetext\vspace{.5cm}
\noindent\rule{.48\linewidth}{.3mm}\rule{.3mm}{.5cm}\vspace{.5cm}}
\def\breakoff{\vspace{.5cm}
\noindent
\rule{.52\linewidth}{.0mm}\rule[-.47cm]{.3mm}{.5cm}\rule{.48\linewidth}{.3mm}
\vspace{.5cm}
\begin{multicols}{2}
\narrowtext}
\documentclass[aps,twocolumn,groupaddress,eqsecnum]{revtex4-1}
\usepackage{graphicx,color}
\usepackage{verbatim}
\usepackage{amssymb}   
\usepackage{amsmath}
\usepackage{amsfonts}
\usepackage{mathdots}
\usepackage{hyperref}
\usepackage{epsfig}

\usepackage{braket}
\usepackage{bm}
\usepackage{physics}

\begin{document}
\title{Topological Fermi-arc-like surface states in Kramers nodal line metals} 
\author{Zi-Ting Sun}
\author{Ruo-Peng Yu}
\author{Xue-Jian Gao}
\author{K. T. Law} 
\thanks{phlaw@ust.hk}

\affiliation{Department of Physics, The Hong Kong University of Science and Technology, Clear Water Bay, Hong Kong, China}

\date{\today} 

\bigskip

\begin{abstract}
The discovery of Kramers nodal line metals (KNLMs) and Kramers Weyl semimetals (KWSs) has significantly expanded the range of metallic topological materials to all noncentrosymmetric crystals. However, a key characteristic of this topology—the presence of topologically protected surface states in KNLMs—is not well understood. In this work, we use a model of a $C_{1v}$ KNLM with curved Kramers nodal lines (KNLs) to demonstrate that Fermi-arc-like surface states (FALSSs), which have a $\mathbb{Z}_2$ topological origin, appear on surfaces parallel to the mirror plane. These states connect two surface momenta, corresponding to the projections of two touching points on the Fermi surfaces. Notably, as achiral symmetries (mirrors and roto-inversions) are gradually broken, the KNLM transitions into a KWS, allowing the FALSSs to evolve continuously into the Fermi arc states of the KWS. We also explore the conditions under which FALSSs emerge in KNLMs with straight KNLs. Through bulk-boundary correspondence, we clarify the topological nature of KNLMs.

\end{abstract}
\bigskip

\maketitle
\section{Introduction}
\label{sec:intro}
In the past two decades, the intersection of topology and symmetry has garnered significant attention in condensed matter physics, particularly following the discovery of time-reversal-invariant topological insulators. These materials are distinguished by an insulating bulk and the presence of topologically protected gapless boundary states \cite{kane2005z,bernevig2006quantum,fu2007topological,zhang2009topological,hasan2010colloquium,qi2011topological}. Researchers quickly recognized that an insulating gap is not a prerequisite for nontrivial topology, leading to the identification of various metallic topological materials, including Dirac semimetals \cite{young2012dirac,wang2012dirac,wang2013three,borisenko2014experimental,liu2014discovery,liu2014stable,yang2014classification,xiong2015evidence,wieder2016double,armitage2018weyl}, Weyl semimetals \cite{armitage2018weyl,wan2011topological,burkov2011weyl,burkov2011topological,xu2011chern,yang2011quantum,halasz2012time,liu2014weyl,hirayama2015weyl,weng2015weyl,huang2015weyl,xu2015discovery,lv2015experimental,soluyanov2015type,ruan2016symmetry}, topological nodal-line semimetals \cite{burkov2011topological,phillips2014tunable,weng2015topological,fang2015topological,mullen2015line,kim2015dirac,yu2015topological,heikkila2015nexus,chan20163,bian2016topological,ezawa2016loop,wang2016body,li2016dirac,yan2016tunable,lim2017pseudospin,hirayama2017topological,behrends2017nodal}, and topological nodal superconductors \cite{sato2010existence,meng2012weyl,yang2014dirac,chiu2014classification,schnyder2015topological,zhao2016unified,huang2018type,zhang2019higher,nayak2021evidence}. Concurrently, the critical role of spatial symmetry in band topology was highlighted with the introduction of topological crystalline insulators \cite{fu2011topological,hsieh2012topological,tanaka2012experimental,dziawa2012topological,xu2012observation,okada2013observation}.

Recently, the classification of metallic topological materials expanded further with the discovery of Kramers Weyl semimetals (KWSs) \cite{chang2018topological} and Kramers nodal line metals (KNLMs) \cite{xie2021kramers}. All nonmagnetic noncentrosymmetric chiral or achiral crystals exhibiting spin-orbit coupling (SOC) are now considered topological, reinforcing the profound connection between symmetry and topology. In KWSs, each two-fold Kramers degeneracy at time-reversal invariant momentum (TRIM) corresponds to a Weyl node with a nonzero topological charge \cite{chang2018topological}, protected by both time-reversal and chiral symmetries. Long Fermi arcs emerge on the surfaces of KWSs, linking the surface projections of Kramers Weyl point pairs (or the projection of pairs of the Fermi pockets) with opposite charges, exemplifying the bulk-surface correspondence. Conversely, in KNLMs, the presence of achiral symmetries (such as mirrors or roto-inversions) transforms band crossing points from discrete zero-dimensional nodes into one-dimensional Kramers nodal lines (KNLs) \cite{xie2021kramers}, which connect TRIMs associated with achiral little groups. KNLMs can be viewed as parent states of KWSs, as the degeneracy of KNLs can be lifted by breaking mirror or roto-inversion symmetries, releasing the Berry flux carried by KNLs. Several recent experiments have confirmed the existence of KNLMs, highlighting phenomena such as unconventional superconductivity in ruthenium silicides \cite{shang2022unconventional}, coexisting KNLs and Weyl fermions in SmAlSi \cite{zhang2023kramers}, and charge density wave-induced KNLs in LaTe$_3$ \cite{sarkar2023charge} and YTe$_3$ \cite{sarkar2024kramers}. More recently, the KNLs have also been identified by ARPES and quantum oscillations in YAuGe~\cite{kurumaji2025electronic} and transition metal dichalcogenides TaS$_2$ and NbS$_2$~\cite{zhang2025kramers,domaine2025tunable}. However, the existence of stable topologically protected surface states in KNLMs remains an open question, representing a crucial aspect of their topological nature.

While the Fermi arcs of KWSs exhibit characteristics similar to those of ordinary Weyl semimetals, in this work, we demonstrate that the topological surface states of KNLMs differ significantly from those found in conventional nodal line semimetals. In the latter, the 2D drumhead surface states are recognized as a defining topological feature~\cite{bian2016drumhead, chan20163, belopolski2019discovery}. In contrast, KNLMs uniquely host topologically protected Fermi-arc-like surface states (FALSSs), which set them apart. To illustrate the presence of FALSSs, we establish a lattice model for a $C_{1v}$ KNLM featuring curved KNLs located on mirror-invariant planes. Our findings reveal that these states appear on KNLM surfaces parallel to the mirror plane. The projection of KNLs divides the surface Brillouin zone into topologically trivial and non-trivial regions, characterized by $0$ and $\pi$ Zak phases, respectively, as determined by mirror symmetry. Unlike the drumhead surface states observed in ordinary nodal line semimetals, the FALSSs of KNLMs exhibit several unique features: (1) they connect points corresponding to the projections of touching points on octdong Fermi surfaces in bulk KNLMs, resembling Fermi arcs that link Weyl nodes with opposite chiralities; (2) they can evolve into the Fermi arc states of a KWS upon breaking achiral symmetries ($e.g.$, through strain), suggesting that FALSSs can be viewed as parent states of the Fermi arcs of KWSs, despite their distinct topological origins. Furthermore, we explore the FALSSs in the $C_{2v}$ KNLM with straight KNLs and clarify the conditions necessary for their observation. Through this analysis, we elucidate the topological characteristics of KNLMs within the framework of bulk-boundary correspondence, paving the way for the experimental observation of this novel phenomenon.

\section{Modelling of KNLMs}
\label{sec:model}
In this section, we use a lattice model with the inclusion of spin-orbit coupling (SOC) to demonstrate the topological surface states of the KNLMs. Here, the symmetry of the model is chosen to be $C_{1v}$ with preserved mirror $m_z$ and time-reversal symmetry (TRS) $\mathcal{T}$, where all the KNLs are curved lines \cite{xie2021kramers}. We will discuss detailedly in the following sections why curved KNLs would be favorable for the existence of the FALSSs.

\begin{figure}[h]
	\centering
	\includegraphics[width=1.1\linewidth]{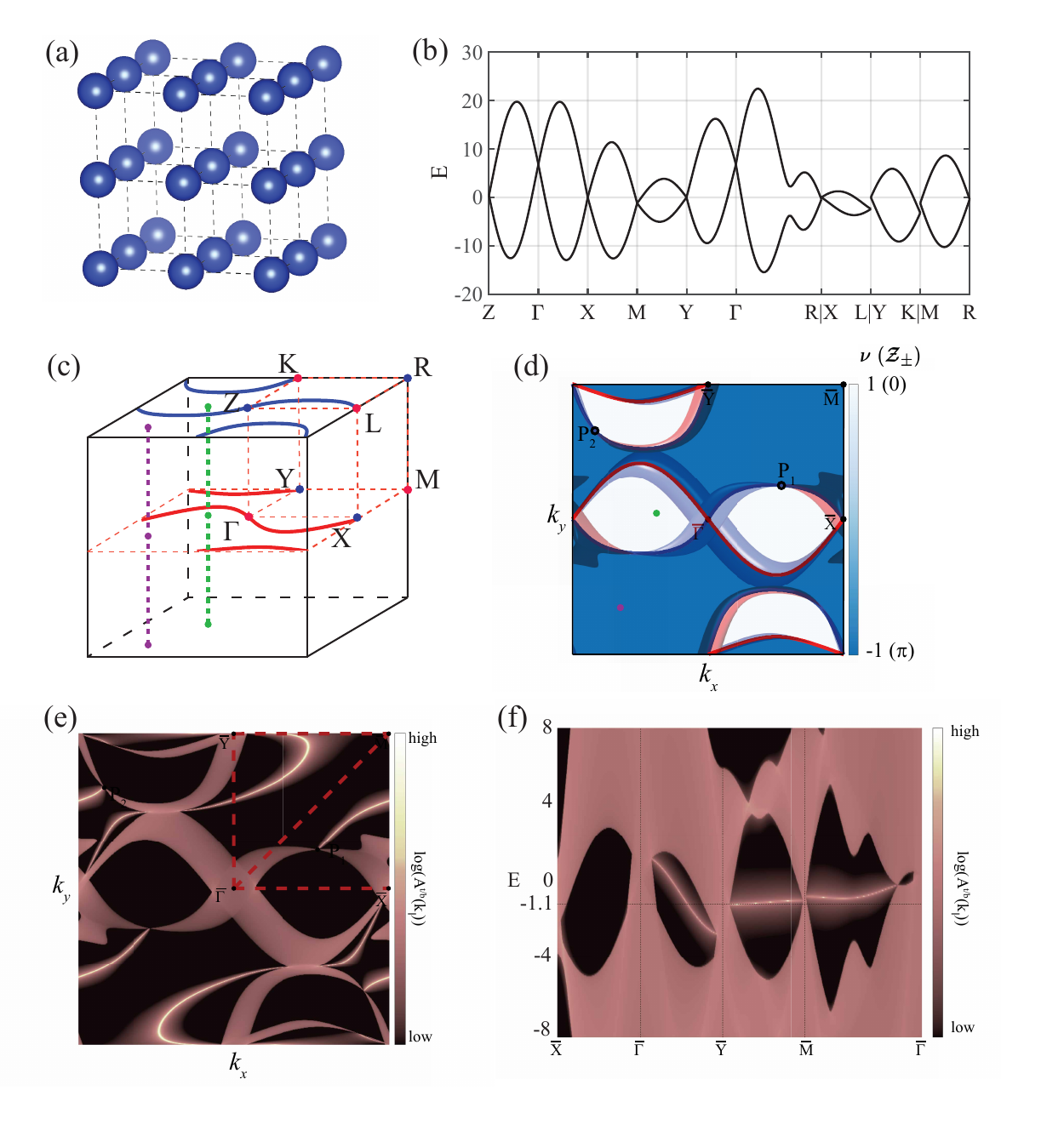}
	\caption{The $C_{1v}$ KLNM. (a) A domatic lattice upon which the $C_{1v}$ tight-binding model is established. (b) The band structure of the Hamiltonian from Eq.~\ref{eq:H_k}. Specific tight-binding parameters are given in Appendix A. (c) The Brillouin zone (BZ) of the $C_{1v}$ model. The KNLs upon the $k_{z}=0$ $(\pi)$ plane are plotted in the red (blue) solid curves. (d) The surface projection of KNLs is the boundary between the topologically non-trivial (cyan) and trivial regions (white). The octdong Fermi surface is also shown with the semi-transparent pink (blue) color representing the electron (hole) pockets. Their touching points, $e.g.$, $P_1$ and $P_2$, are on the KNLs. (e) and (f) show the surface spectral function $A^{t/b}(\bm{k}_\parallel,E)$ on the (001) surface of the $C_{1v}$ tight-binding model. The energy level in (e) is set as $E=-1.1$, as indicated by the horizontal black dashed line in (f).
	}
	\label{fig:fig1}
\end{figure}

For simplicity, the tight-binding Hamiltonian is defined upon a cubic lattice (Fig.\ref{fig:fig1}(a)) as 
\begin{equation}
	\label{eq:H_k}
	\mathcal{H}=\sum_{\bm{R}_{i}}\sum_{\bm{\tau},\mu\nu}H_{\bm{\tau},\mu\nu}c_{\bm{R}_{i},\mu}^{\dagger}c_{\bm{R}_{i}+\bm{\tau},\nu}.
\end{equation}
with hopping matrices $H_{\bm{\tau},\mu\nu}=\langle0,\mu|\hat{H}|\bm{\tau},\nu\rangle$. Here $\mu,\nu=\uparrow,\downarrow$ represents the spin indices and $\bm{\tau}$ is the hopping vector. To break the extra symmetries brought by the cubic lattice, we consider all the $m_z$- and $\mathcal{T}$-invariant hoppings up to the next-nearest-neighbors ($i.e.$, $|\bm{\tau}|\leqslant\sqrt{2}a$ with the lattice constant $a=1$ hereinafter). The concrete form of this tight-binding model and its hopping parameters are given in Appendix A.

The energy spectrum of this Hamiltonian with suitable parameters is shown in Fig.~\ref{fig:fig1}(b). Apart from the Kramers degeneracy at the TRIMs, the achiral $C_{1v}$ symmetry further requires line nodes, $i.e.$, the KNLs, to exist upon the mirror-invariant planes where $k_{z}=0$ or $\pi$, as shown by the red and blue solid curves in Fig.~\ref{fig:fig1}(c,d). This can be easily understood by the following argument. 
We first denote the SOC part of the total Hamiltonian as $H_{soc}=\bm{d}(\bm{k})\cdot \bm{\sigma}$.
Upon the mirror-invariant planes, the mirror $m_{z}$ symmetry
requires that only the nonzero component of $\bm{d}(\bm{k})$ is
the z-component, $i.e.$, the spins of the eigenstates are polarized along
the z-direction, which is similar to the case of the 2D Ising superconductors \cite{lu2015ising}.
The TRS dictates the oddity of function $d^{z}(\bm{k}_\parallel,k_z=0(\pi))$ with $d^{z}(\bm{k}_\parallel,k_z = 0(\pi))=-d^{z}(-\bm{k}_\parallel,k_z=0(\pi))$, and the continuity of the function assures the existence of nodal lines indicated by the equation $d^{z}(\bm{k}_\parallel,k_z=0(\pi))=0$, which lie within the mirror-invariant planes and connect different TRIMs. Here $\bm{k}_\parallel = (k_x, k_y)$. These nodal lines are named as Kramers nodal lines (KNLs) and the corresponding materials as Kramers nodal line metals (KNLMs), which have been studied in Ref.~\cite{xie2021kramers}. 

However, one essential part to claim that KNLMs are topological materials -- the bulk-boundary correspondence has been missed in the previous work. Therefore, in the following sections, we intend to fill in this important part by studying the stable surface states that are protected by the topology of KNLs.

\section{FALSSs induced by the KNLs}
\label{sec:surface}

To show the surface states of KNLMs, we use the slab geometry with the normal vector of the surfaces along the z-direction, which is also perpendicular to the mirror plane. The surface spectral function is calculated with $A^{t/b}(\bm{k}_\parallel,E)=-\frac{1}{\pi}\Im\Tr G^{r,t/b}(\bm{k}_\parallel,E)$ where $G^{r,t/b}(\bm{k}_\parallel,E)$ is the retarded Green's function of the top/bottom surface, and the results are shown in Fig.~\ref{fig:fig1}(e,f). The mirror symmetry relates the two surfaces as $A^{t}(\bm{k}_\parallel,E)=A^{b}(\bm{k}_\parallel,E)$. Interestingly, we find that FALSSs are appearing on both surfaces of the slab, with each arc connecting two points ($P_1$ and $P_2$ in Fig.~\ref{fig:fig1}(d)) which are both the projection of the touching points of the so-called octdong Fermi surface \cite{xie2021kramers} in bulk KNLMs. This is in contrast with the Fermi arc states in Weyl semimetals, where the Fermi arcs connect the projection of a pair of Weyl nodes. In the following of this section, we will study in detail the origin and the exotic properties of these FALSSs induced by the KNLs.

Being a gapless metallic system, it is impossible to define a topological invariant over the whole Brillouin zone, but this can be solved by reducing the dimension of the system. Particularly for KNLMs, we can set $\bm{k}_\parallel = (k_x, k_y)$ as new parameters and study the topology of the 1D Hamiltonian $H_{\bm{k}_\parallel}(k_z)$. We demonstrate here in a pedagogical way that the Zak phase of the 1D Hamiltonian $H_{\bm{k}_\parallel}(k_z)$ with half-integer spin is quantized if the system is invariant under the mirror $m_z$, and the topological invariant is of $\mathbb{Z}_2$-type. A more general proof is included in Appendix C. We need to point out here that similar conclusions are drawn in Ref.~\cite{chan20163} in a system with negligible SOC. 
Interestingly, both cases are not included in the classification of mirror-symmetry-protected topological semimetals \cite{chiu2014classification}. Moreover, our case is different from the 1D topological mirror insulator in Ref.~\cite{lau2016topological}, which is protected by both mirror symmetry and TRS. In contrast, TRS is not preserved in the reduced 1D Hamiltonian $H_{\bm{k}_\parallel}(k_z)$ of KNLMs.

\begin{figure}
	\centering
	\includegraphics[width=1\linewidth]{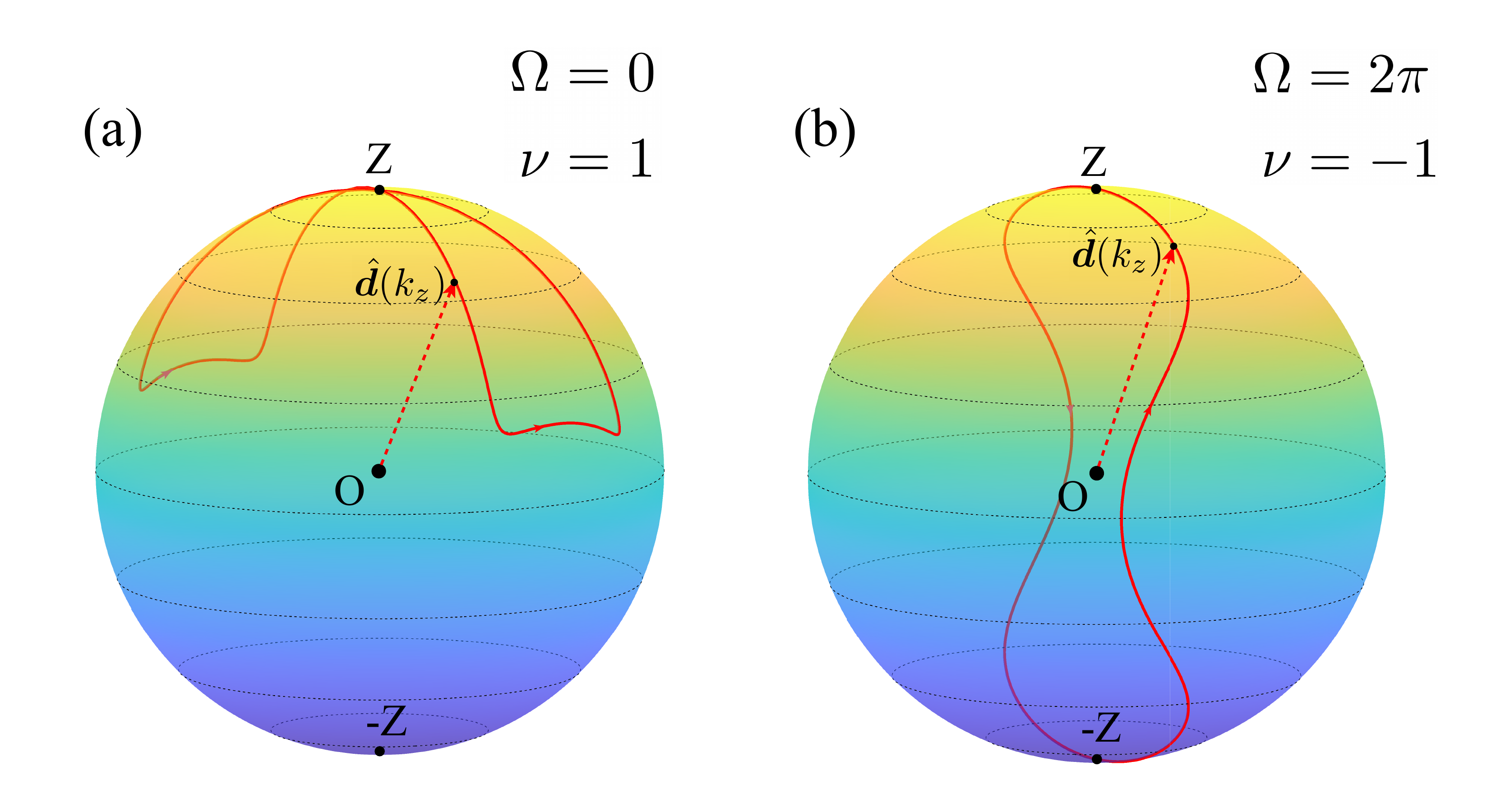}
	\caption{ The $\mathbb{Z}_2$ topological invariant protected by the mirror symmetry.
	(a) and (b) show how the spin rotates upon the unit sphere when $k_z$ of the 1D Hamitonian $H_{\bm{k}_\parallel}(k_z)$ changes from $-\pi$ to $\pi$, respectively for topologically trivial ($\mathcal{Z}=0$) and non-trivial case ($\mathcal{Z}=\pi$).
	}
	\label{fig:fig2}
\end{figure}

Suppose we are studying a two-band model with spin-orbit coupling, and the Zak phase for the reduced 1D Hamiltonian $H_{\bm{k}_\parallel}(k_z)$ can be formulated as 
\begin{align}
	\mathcal{Z}_\pm (\bm{k}_\parallel) & = \int_{-\pi}^{\pi}A^z_{\pm}(\bm{k})\mathrm{d}k_z \mod 2\pi \nonumber\\
	& = \frac{\Omega_{\pm}(\bm{k}_\parallel)}{2} \mod 2\pi.
	\label{Zak_phase}
\end{align}
In the first line, $\pm$ represents the upper/lower band, $A^z_{\pm}(\bm{k})=i\langle\phi_{\pm \bm{k}}|\partial_{k_z}|\phi_{\pm \bm{k}}\rangle$ is the z-component of the Berry connection, and the use of ``mod $2\pi$" is due to the gauge variance of $\mathcal{Z}_\pm (\bm{k}_\parallel)$ by $2n\pi$. 
More importantly, the second line of Eq.~\ref{Zak_phase} points out the well-known geometrical meaning of $\mathcal{Z}_\pm (\bm{k}_\parallel)$, where $\Omega_{\pm}(\bm{k}_\parallel) = \frac{1}{2}\oint_{\mathcal{C}_{\pm}(\bm{k}_\parallel)}\mathrm{d}S$ is the surface area on a unit sphere bounded by the loop $\mathcal{C}_{\pm}(\bm{k}_\parallel)$, which is the path swept by the spin direction $\bm{s}_\pm=\langle\phi_{\pm}|\bm{\sigma}|\phi_{\pm}\rangle = \pm\bm{d}/|\bm{d}|=\pm\hat{\bm{d}}$ of an upper/lower band electron of a fixed $\bm{k}_\parallel$ but with its $k_z$ varying from $-\pi$ to $\pi$ (as shown in Fig.~\ref{fig:fig2}).

Now let us study the role that the mirror symmetry $m_z$ plays here. $m_z$ requires that 
\begin{align}
	d^{x,y}(\bm{k}_\parallel,-k_z)&=-d^{x,y}(\bm{k}_\parallel,k_z), \\
	d^{z}(\bm{k}_\parallel,-k_z)&=\ d^{z}(\bm{k}_\parallel,k_z),
\end{align}
which, graphically, means each pair of points $\hat{\bm{d}}(k_z)$ and $\hat{\bm{d}}(-k_z)$ upon the loop $\mathcal{C}_{\pm}(\bm{k}_\parallel)$ with $k_z$ varying from 0 to $\pi$ will bisect the circle of constant $\hat{d}^z=\hat{d}^z(\pm k_z)$ (thin black dashed circles in Fig.~\ref{fig:fig2}) on the unit sphere. 
Especially we have $d^{x,y}(k_z = 0, \pi)=0$, in other words $\bm{d}(k_z = 0, \pi)\parallel \hat{\bm{z}}$. Depending on the value of $\nu=\mathrm{sgn} [d^{z}(k_z=0)\cdot d^{z}(k_z=\pi)]$, the spin loop $C_{\pm}(\bm{k}_\parallel)$ shows two distinct scenarios, and this 1D system $H_{\bm{k}_\parallel}(k_z)$ can therefore be classified into two topologically distinct cases. For the $\nu=1$ case, the spin loop passes the same pole of the spin sphere twice at $k_z=0$ and $\pi$, resulting in a topologically trivial phase with $\mathcal{Z}_{\pm}=0$ (Fig.~\ref{fig:fig2}(a)). In comparison for the $\nu=-1$ cases, the spin loop passes both poles and bisects the unit sphere surface, giving rise to a topologically non-trivial phase with $\mathcal{Z}_{\pm}=\pi$ (Fig.\ref{fig:fig2}(b)). In short, the above argument can be summarized into a single formula as
\begin{align}
	\mathcal{Z}_{\pm}(\bm{k}_\parallel) & = \frac{\pi}{2}[1-\nu(\bm{k}_\parallel)] \nonumber \\
		 & = \frac{\pi}{2}[ 1-\mathrm{sgn} \left( d^z(\bm{k}_\parallel,0)\cdot d^z(\bm{k}_\parallel,\pi) \right) ].
\end{align}

This quantized Zak phase will not change as long as the reduced 1D system $H_{\bm{k}_\parallel}(k_z)$ remains fully gapped. It is then natural to see that the surface projections of KNLs mark the boundaries of the topological trivial and non-trivial regions (Fig.~\ref{fig:fig1}(d)), as the 1D system becomes gapless when $\bm{k}_{\parallel}$ goes across these projection lines. The $\pi$ winding phase of these KNLs \cite{xie2021kramers} further assures their role as the boundaries of different topological regions. More interestingly, the projection of KNLs marks the ending of the arc-like surface states, and the two ending points are exactly the projection of the octdong Fermi surface touching points ($P_1$ and $P_2$ in Fig.~\ref{fig:fig1}(e)). By gradually tuning the chemical potential, the endpoints of these arc states will move along the projection lines of KNLs and the arcs will sweep across the whole topological region that is edged by the projection of KNLs.

\section{Conversion of the FALSSs of KNLMs to the Fermi arcs of KWSs}
\label{sec:conver}
In the sense of symmetry and bulk spectrum properties, KNLMs can be viewed as the parent state of the KWSs, as pointed out in Ref.~\cite{xie2021kramers}. When all the mirrors and roto-inversions of the originally achiral space group symmetries are broken (by, $e.g.$, strain), the whole system will become chiral. The KNLs, which are protected by the achiral space group symmetry and TRS, will then be gapped out, and only the Kramers degeneracies at the TRIMs will survive due to the still preserved TRS, forming the Kramers Weyl nodes. 

Concerning the FALSSs of the KNLMs, a natural question to ask is whether the FALSSs of a KNLM will disappear or not when an achiral KNLM becomes a chiral KWS with KNLs being gapped out. In this section, we point out that these FALSSs will persist in this process and evolve into the Fermi arc states of the KWS. In this sense, the FALSSs of KNLMs can also be regarded as the parent states of the real Fermi arc states of KWSs.

\begin{figure}
	\centering
	\includegraphics[width=1\linewidth]{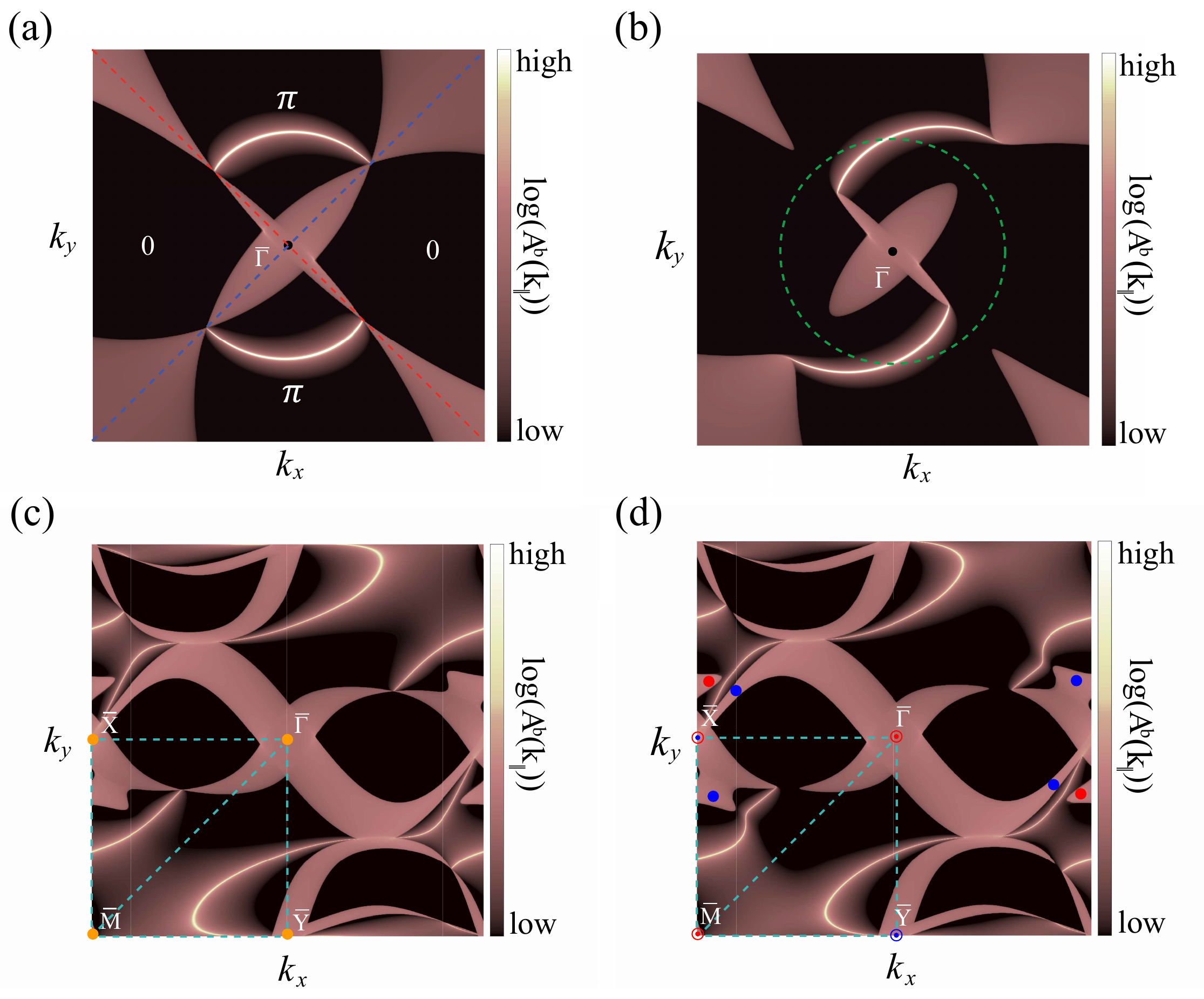}
	\caption{ The FALSSs of the KNLM evolve into the Fermi arc states of the KWS.
	(a) shows the (001)-surface spectral weight $A(\bm{k}_\parallel,E=0)$ of the $C_{1v}$ KNLM $k\cdot p$ Hamiltonian $h(\bm{k})$ of Eq.~\ref{eq:kp_model} with $m=0.2, t_z=0.02, \mu=0.2, \alpha=1.5, \beta=1, a_1=4, a_2=2, b_1=2, b_2=4$. The red (blue) dashed line is the projection of the KNL in the $k_z=0\,(\pi)$ plane, which separates the topologically trivial and nontrivial regions with 0 and $\pi$ Zak phase, respectively.
	(b) shows the (001)-surface spectral weight when the mirror-breaking term $h^\prime(\bm{k})$ from Eq.~\ref{eq:mirror_breaking} with $c=1$ is added. Now the chirality of Kramers Weyl nodes at $\Gamma$ and Z are both negative, and the Chern number defined on the torus surface whose cross-section with constant $k_z$ planes is indicated by the green circle is -2. (c) and (d) show the (001)-surface spectral function $A(\bm{k}_{\parallel}, E=-1.1)$ of the lattice models, respectively for the $C_{1v}$ KNLM and the $C_1$ KWS. The parameters of the mirror-breaking term in Eq.~\ref{eq:mirror-breaking} are set as $\beta_1=0.2, \beta_2=-0.1$. In (d), the surface projections of the Weyl nodes with chirality +1(-1) are labeled by red(blue) dots. For the Kramers Weyl nodes at the TRIMs, the chirality of the Weyl node on the $k_z=0(\pi)$ plane is denoted by the outer (inner) circles.
	}
	\label{fig:fig3}
\end{figure}

For the sake of analysis, yet without loss of generality, we can expand the Hamiltonian around the $\Gamma$-Z line to get a $k\cdot p$ model as
\begin{equation}
	h(\bm{k})=\varepsilon_{0}(\bm{k})\sigma_{0}+\bm{d}(\bm{k})\cdot\bm{\sigma}
	\label{eq:kp_model}
\end{equation}
with
\begin{equation}
	\varepsilon_0(\bm{k}) = \frac{k_x^2+k_y^2}{2m}+t_z\cos k_z - \mu,
\end{equation}
and
\begin{multline}
	\bm{d}(\bm{k}) = \bm{(} \alpha \sin k_z,  \beta \sin k_z,\\
	(a_1+a_2\cos k_z)k_x + (b_1+b_2\cos k_z)k_y \bm{)}.
\end{multline}
With much fewer parameters, this model captures the main features of the KNLMs and their surface states. As shown in Fig.~\ref{fig:fig3}(a), when projected onto the (001) surface, the projections of two KNLs, which originally lie on the $k_z=0$ and $\pi$ planes, separate the surface Brillouin zone into topologically trivial and nontrivial regions. The FALSSs, lying in the topological nontrivial region with Zak phase $\pi$, connect the projection of the octdong Fermi surface touching points. 

We can now introduce an extra term which breaks mirror $m_z$ but preserves $\mathcal{T}$ 
\begin{equation}
	h^\prime(\bm{k}) = \gamma k_x \sigma_x,
	\label{eq:mirror_breaking}
\end{equation}
which can gap out all the KNLs and make the $C_{1v}$ KNLM into a simple $C_1$ KWS. The chirality of two Kramers Weyl node at $\Gamma$ and Z are $C_\Gamma = -\mathrm{sgn}[\gamma\beta(b_1+b_2)]$ and $C_Z = \mathrm{sgn}[\gamma\beta(b_1-b_2)]$ respectively. When $C_\Gamma$ and $C_Z$ are both positive or negative, the FALSSs of KNLMs will evolve into the Fermi arc states of the KWS, as shown in Fig.~\ref{fig:fig3}(b). Unlike the FALSSs of KNLMs whose topological origin is the mirror-protected $\pi$ Zak phase, here the Fermi arc states are due to the nonzero Chern number defined on the torus surface whose cross-section with constant $k_z$ planes is indicated by
the green circle in Fig.~\ref{fig:fig3}(b). A similar evolution can be realized in the lattice model as well. When an extra term $H^{\prime}({\bm{k}})$ which breaks the mirror symmetry but preserves the TRS
		\begin{equation}
			H^{\prime}({\bm{k}}) = \beta_1 \sin(k_x) \sigma_x  + \beta_2 \sin(k_y) \sigma_y.
			\label{eq:mirror-breaking}
		\end{equation}
		is added to the original Hamiltonian $H(\bm{k})$, the $C_{1v}$ KNLM will now become a $C_1$ KWS, with doubly-degeneracy only surviving on the TRIMs, forming the Kramers Weyl nodes. Correspondingly, the FALSSs will continuously evolve into the Fermi arcs of the KWS, as shown in Fig.~\ref{fig:fig3}(c) and (d). The coevolution of bulk and boundary, from KNLMs to KWSs, illustrates the principle of bulk-edge correspondence in topological materials.

\section{FALSSs of straight KNLs}
\label{sec:straight}
In the previous sections, we have shown that the projection of curved KNLs upon the surface that is parallel to the mirror plane ($e.g.$, (001) surface in the $m_z$ case) marks the boundary between the topologically trivial and non-trivial regions within the surface Brillouin zone. This scenario is guaranteed by three facts: (1) the mirror symmetry $m_z$ forces the Zak phase to be quantized (0 or $\pi$) for a quasi-1D Hamiltonian $H_{K_x,k_y}(k_z)$; (2) the winding Berry phase of a linear-dispersed KNL is $\pi$; (3) the projection of KNLs from $k_z=0$ plane and from $k_z=\pi$ plane, in general, do not overlap with each other.

\begin{figure}
	\centering
	\includegraphics[width=1\linewidth]{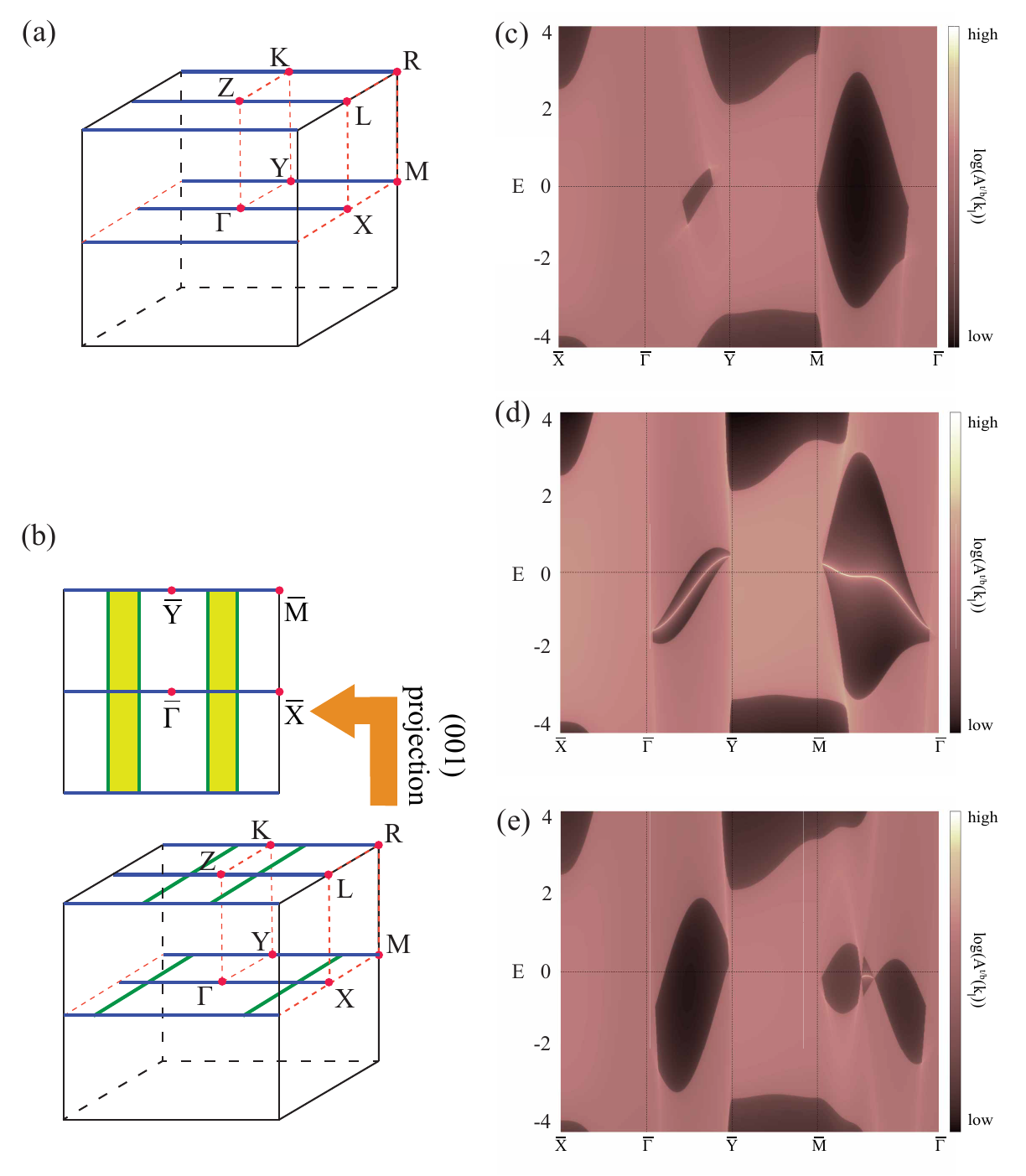}
\caption{The $C_{2v}$ KNLM. 
			(a,b) show the KNLs (blue) and the extra nodal lines (green) in the first Brillouin zone. (c,d,e) show the surface spectrum under three different parameter regimes. When there is no extra nodal line other than the KNLs, the whole (001)-surface Brillouin zone is either completely trivial (as shown in (c)) or completely topological (as shown in (d)). When the extra nodal lines are present, their surface projection will cut the (001)-surface Brillouin zone into trivial and topological (yellow areas in (b)) regions. 
			}
	\label{fig:fig4}
\end{figure}

However, if extra symmetries other than a single mirror are imposed, $e.g.$, the 2, 3, 4, 6-fold rotations or another mirror, the originally curved KNLs will be pinned along a high-symmetry line (Fig.~\ref{fig:fig4}(a)) and then the condition (3) mentioned above will no longer hold. For a surface that is parallel to the mirror plane, when a surface $\bm{k}_\parallel$-vector moves across a line that is the overlapping projection of two KNLs each with a $\pi$ winding phase, the Zak phase of the corresponding quasi-1D Hamiltonian will not change as in the $C_{1v}$ case. Thus, the projection of KNLs is no longer the boundary of topologically trivial and non-trivial regions, and the existence of topologically nontrivial regions in the surface Brillouin zone is not guaranteed in this case. Nevertheless, the surface states are still possible to exist in this kind of KNLMs, when the whole surface Brillouin zone is topologically non-trivial. In the following of this section, we will introduce a $C_{2v}$ KNLM model as an example.

Besides $m_z$, another generator of the $C_{2v}$ point group with elements $\{E,C_{2x},m_y,m_z\}$ can be chosen as $m_y$. Correspondingly, besides the constraints from TRS and $m_z$, $m_y=-i\sigma_y$ imposes an additional requirement. We still adopt the same cubic lattice as in the former section and consider the hoppings up to the third-nearest-neighbor with mutually independent terms. The details of the model are given in Appendix B. There are three topologically distinct parameter regions for this tight-binding model considering the existence of Fermi-arc-like states on the (001) surface. (1) The whole (001) surface is topologically trivial, as shown in Fig.~\ref{fig:fig4}(c); (2) The whole (001) surface is topologically non-trivial, as shown in Fig.~\ref{fig:fig4}(d). Thus the observable FALSSs can emerge; (3) When there are extra nodal lines (besides the KNLs) lying upon the $k_z=0$ or $k_z=\pi$ planes (Fig.~\ref{fig:fig4}(b)), the projection of which will cut the (001) surface Brillouin zone into topologically trivial and non-trivial regions, as shown in Fig.~\ref{fig:fig4}(e). The emergence of these extra nodal lines is due to the sign change of $d^z(\bm{k})$ upon the mirror-invariant plane. Unlike the KNLs, these extra nodal lines are not symmetrically protected and can be annihilated in pairs.

\section{Discussion}
\label{sec:concl}

Here we discuss a little bit about the requirements of the band dispersion for the existence of observable FALSSs in a real KNLM material. Note that the topological origin of the FALSSs of KNLMs is the quantized $\pi$ Zak phase of the quasi-1D subsystem $H_{\bm{k}_\parallel}(k_z)$ where $k_z$ is perpendicular to the mirror plane (Fig.~\ref{fig:fig1}(c)). A positive-definite gap opened by SOC (rather than an indirect negative gap) of this quasi-1D subsystem is required for the stability of its end states, $i.e.$, the FALSSs of KNLMs. Otherwise, the surface states will merge into the bulk. This condition is actually equivalent to the requirement of the octdong-type Fermi surface (rather than the spindle-torus Fermi surface) of KNLMs, which appear in the presence of a sizable SOC strength and relatively flat bands. Recently, KNLs with tunable octdong Fermi surfaces have been reported in KNLM 3R-TaS$_2$~\cite{domaine2025tunable}. At the same time, the KNL connecting TRIMs $L$ and $F$ is free to disperse on the mirror plane, which is similar to the case of curved KNLs we considered above. So we expect this material to provide a promising platform to observe FALSSs we reported in this paper.

\begin{acknowledgments} 
We thank Ying-Ming Xie for inspiring discussions. K. T. L. acknowledges the support of the Ministry of Science and Technology, China, and Hong Kong Research Grant Council through Grants No. 2020YFA0309600, No. RFS2021-6S03, No. C6025-19G, No. AoE/P-701/20, No. 16310520, No. 16307622, and No. 16309223.
\end{acknowledgments}

\appendix 
 
\section{Lattice model of $C_{1v}$ KNLM} 
\label{app:c1v} 
In the main text, we use a tight-binding model of a $C_{1v}$ KNLM to exemplify the surface states of KNLMs. Here we show in detail how it is derived.

		The symmetry respected in this system is the TRS $\mathcal{T}$ and a mirror $m_z$.
		We denote the hopping matrix as $H_{\bm{\tau},\mu\nu}=\langle0,\mu|\hat{H}|\bm{\tau},\nu\rangle$ with $\mu,\nu=\uparrow,\downarrow$ representing the spin indices and $\bm{\tau}$ the hopping vector, then the Hamiltonian in the real space can be written as
		\begin{equation}
			\hat{H}=\sum_{\bm{R}_{i}}\sum_{\bm{\tau},\mu\nu}H_{\bm{\tau},\mu\nu}c_{\bm{R}_{i},\mu}^{\dagger}c_{\bm{R}_{i}+\bm{\tau},\nu}.
		\end{equation}
		The TRS $\hat{\mathcal{T}}=i\sigma_y \hat{\mathcal{K}}$ requires that 
		\begin{equation}
		H_{\bm{\tau}}=\sigma_{y}H_{\bm{\tau}}^*\sigma_{y},
			\label{eq:T_require}
		\end{equation}
		the horizontal mirror $m_z=-i\sigma_z$ requires that 
		\begin{equation}
			H_{\bm{\tau}}=\sigma_{z}H_{{m_z}\bm{\tau}}\sigma_{z},
			\label{eq:mz_require}
		\end{equation}
		and the periodicity of the lattice requires that 
		\begin{equation}
			H_{-\bm{\tau}}=H_{\bm{\tau}}^{\dagger}.
			\label{eq:periodic_require}
		\end{equation}
		From Eq.~\ref{eq:mz_require}, specifically we have
		\begin{align}
			H_{\bm{\tau}} & =\left(\begin{array}{cc}
			\alpha_{1,\bm{\tau}} & 0\\
			0 & \alpha_{1,\bm{\tau}}^{*}
			\end{array}\right),\ \mathrm{for\ }\bm{\tau}\perp\hat{\bm{z}},\\
			H_{\bm{\tau}} & =\left(\begin{array}{cc}
			\mathrm{Re}(\alpha_{1,\bm{\tau}}) & \alpha_{2}\\
			-\alpha_{2}^{*} & \mathrm{Re}(\alpha_{1,\bm{\tau}})
			\end{array}\right),\ \mathrm{for\ }\bm{\tau}\parallel\hat{\bm{z}}.
		\end{align}
		By Fourier-transforming into the reciprocal space, the Hamiltonian
		can be written as
		\begin{equation}
			\hat{H}=\sum_{\bm{k},\mu\nu}\left[H(\bm{k})\right]_{\mu\nu}c_{\bm{k}\mu}^{\dagger}c_{\bm{k}\nu}
		\end{equation}
		with $H(\bm{k})=\sum_{\bm{\tau}}H_{\bm{\tau}}e^{i\bm{k}\cdot\bm{\tau}}.$

		For simplicity, we adopt a cubic lattice (as shown in the main text Fig.~1(a)) and break the extra symmetries (other than a mirror) by anisotropic hoppings. The lattice constant is set to be 1 hereinafter.
		As shown in the main text, we consider the hoppings up to the second-nearest-neighbor, explicitly with mutually independent hoppings as
		\begin{align}
			H_{(100)} &= t_{x}\sigma_{0}+i\alpha_{x}\sigma_{z}, \\
			H_{(010)} &= t_{y}\sigma_{0}+i\alpha_{y}\sigma_{z}, \\
			H_{(001)} &= t_{z}\sigma_{0}+i\alpha_{z}^x\sigma_{x}+i\alpha_{z}^y\sigma_{y}, \\
			H_{(110)} &= t_{xy}\sigma_{0}+i\alpha_{xy}\sigma_{z}, \\
			H_{(1\bar{1}0)} &= t_{x\bar{y}}\sigma_{0}+i\alpha_{x\bar{y}}\sigma_{z}, \\
			H_{(101)} &= t_{zx}\sigma_{0}+i\sum_{j=x,y,z}\alpha_{zx}^j\sigma_{j}, \\
			H_{(011)} &= t_{zy}\sigma_{0}+i\sum_{j=x,y,z}\alpha_{zy}^j\sigma_{j}.
		\end{align}
		Written in the reciprocal space, the Hamiltonian is 
		\begin{equation}
			H(\bm{k})=\epsilon_{0}(\bm{k})\sigma_{0}+\bm{d}(\bm{k})\cdot\boldsymbol{\sigma}
		\end{equation}
		with
        \begin{equation}
            \begin{aligned}
			\epsilon_{0}(\bm{k})&=\sum_{i=x,y,z}2t_{i}\cos k_{i}+2t_{xy}\cos(k_{x}+k_{y})-\mu \\
            &+2t_{x\bar{y}}\cos(k_{x}-k_{y})+4(t_{zx}\cos k_{x}+t_{zy}\cos k_{y})\cos k_{z},
		\end{aligned}
        \end{equation}
		\begin{equation}
			d^{x}(\bm{k})=-2(\alpha_{z}^x+2\alpha_{zx}^{x}\cos k_{x}+2\alpha_{zy}^{x}\cos k_{y})\sin k_{z},
		\end{equation}
		\begin{equation}
			d^{y}(\bm{k})=-2(\alpha_{z}^y+2\alpha_{zx}^{y}\cos k_{x}+2\alpha_{zy}^{y}\cos k_{y})\sin k_{z},
		\end{equation}
		\begin{multline}
			d^{z}(\bm{k})=-2[\alpha_{x}\sin k_{x}+\alpha_{y}\sin k_{y}+\alpha_{xy}\sin(k_{x}+k_{y}) \\ +\alpha_{x\bar{y}}\sin(k_{x}-k_{y})+2(\alpha_{zx}^{z}\sin k_{x}+\alpha_{zy}^{z}\sin k_{y})\cos k_{z}].
		\end{multline}

		In the main text, we use the following set of parameters throughout the tight-binding calculations for this model: $t_{x}=0.5,t_{y}=0.6,t_{xy}=0.3,t_{x\bar{y}}=0.4,t_{z}=0.7,t_{zx}=0.25,t_{zy}=0.2,\mu=0;$ $\alpha_{x}=2,\alpha_{y}=2.4,\alpha_{xy}=1.5,\alpha_{x\bar{y}}=1.4,\alpha_{z}^y=1.9,\alpha_{z}^x=0.9,\alpha_{zx}^{x}=1.35,\alpha_{zx}^{y}=1.1,\alpha_{zx}^{z}=1.55,\alpha_{zy}^{x}=1.43,\alpha_{zy}^{y}=0.25,\alpha_{zy}^{z}=1.85.$

\section{Lattice model of $C_{2v}$ KNLM} 
\label{app:c2v} 
In this section, we show the details of the model describing a $C_{2v}$ KNLM model. Besides $m_z$, another generator of the $C_{2v}$ point group with elements $\{E,C_{2x},m_y,m_z\}$ can be chosen as $m_y$. Correspondingly, besides the requirements from Eq.~\ref{eq:T_require} to Eq.~\ref{eq:periodic_require}, $m_y=-i\sigma_y$ imposes an additional requirement 
		\begin{equation}
H_{\bm{\tau}}=\sigma_{y}H_{\hat{m}_{y}\bm{\tau}}\sigma_{y}.
		\end{equation}

		We still adopt the same cubic lattice as in the former section and consider the hoppings up to the third-nearest-neighbor with mutually independent terms
		\begin{align}
			H_{(100)} &= t_{x}\sigma_{0}, \\
			H_{(010)} &= t_{y}\sigma_{0}+i\alpha_{y}^z\sigma_{z}, \\
			H_{(001)} &= t_{z}\sigma_{0}+i\alpha_{z}^y\sigma_{y}, \\
			H_{(011)} &= t_{yz}\sigma_{0}+i\alpha_{yz}^y\sigma_{y} + i\alpha_{yz}^z\sigma_{z}, \\
			H_{(101)} &= t_{zx}\sigma_{0}+i\alpha_{zx}^y\sigma_{y}, \\
			H_{(110)} &= t_{xy}\sigma_{0}+i\alpha_{xy}^z\sigma_{z}, \\
			H_{(111)} &= t_{xyz}\sigma_{0}+i\sum_{j=x,y,z}\alpha_{xyz}^j\sigma_{j}.
		\end{align}
		Written in the reciprocal space, the Hamiltonian is 
		\begin{equation}
			H(\bm{k})=\epsilon_{0}(\bm{k})\sigma_{0}+\bm{d}(\bm{k})\cdot\boldsymbol{\sigma}
		\end{equation}
		with
        \begin{equation}
            \begin{aligned}
			\epsilon_{0}(\bm{k})&=\sum_{i=x,y,z}2t_{i}\cos k_{i}+\sum_{ij=xy,yz,zx}4t_{ij}\cos k_{i}\cos k_{j}\\
            &+8t_{xyz}\cos k_x \cos k_y \cos k_z - \mu,
        \end{aligned}
        \end{equation}        
		\begin{equation}
			d^{x}(\bm{k})=8\alpha_{xyz}^x \sin k_x \sin k_y \sin k_z,
		\end{equation}
		\begin{multline}
			d^{y}(\bm{k})=-2\alpha_z^y \sin k_z - 4 \alpha_{yz}^y \cos k_y \sin k_z \\- 4 \alpha_{zx}^y \cos k_x \sin k_z - 8 \alpha_{xyz}^y \cos k_x \cos k_y \sin k_z,
		\end{multline}
		\begin{multline}
			d^{z}(\bm{k})=-2\alpha_y^z \sin k_y + 4\alpha_{yz}^z\sin k_y \cos k_z \\-4\alpha_{xy}^z\cos k_x \sin k_y - 8\alpha_{xyz}^z\cos k_x \sin k_y \cos k_z.
		\end{multline}

		There are three topologically distinct parameter regions for this tight-binding model considering the existence of Fermi-arc-like states on the (001) surface:

		(a) When $(\alpha_y^z-2\alpha_{yz}^z)^2 > 4(\alpha_{xy}^z+2\alpha_{xyz}^z)^2$, $(\alpha_y^z+2\alpha_{yz}^z)^2 > 4(\alpha_{xy}^z-2\alpha_{xyz}^z)^2$ and $(\alpha_y^z-2\alpha_{yz}^z)(\alpha_y^z+2\alpha_{yz}^z)>0$ are all satisfied, the whole (001) surface is topologically trivial. Corresponding to Fig.~\ref{fig:fig4}(c) in the main text.

		(b) When $(\alpha_y^z-2\alpha_{yz}^z)^2 > 4(\alpha_{xy}^z+2\alpha_{xyz}^z)^2$, $(\alpha_y^z+2\alpha_{yz}^z)^2 > 4(\alpha_{xy}^z-2\alpha_{xyz}^z)^2$ and $(\alpha_y^z-2\alpha_{yz}^z)(\alpha_y^z+2\alpha_{yz}^z)<0$ are all satisfied, the whole (001) surface is topologically non-trivial. Corresponding to Fig.~\ref{fig:fig4}(d) in the main text.

		(c) When either $(\alpha_y^z-2\alpha_{yz}^z)^2 < 4(\alpha_{xy}^z+2\alpha_{xyz}^z)^2$ or $(\alpha_y^z+2\alpha_{yz}^z)^2 < 4(\alpha_{xy}^z-2\alpha_{xyz}^z)^2$ is satisfied, there will be the extra nodal lines (besides the KNLs) lying upon the $k_z=0$ or $k_z=\pi$ planes, the projection of which will cut the (001) surface Brillouin zone into topologically trivial and non-trivial regions. Corresponding to Fig.~\ref{fig:fig4}(e) in the main text.

The tight-binding parameters for plotting Fig.~\ref{fig:fig4}(c,d,e) are set as $\mu=0,t_z=0.11,t_y=0.12,t_{yz}=0.13,t_x=0.14,t_{xy}=0.15,t_{xz}=0.16,t_{xyz}=0.17;\alpha_z^y=1.9,\alpha_{yz}^y=0.25,\alpha_{xy}^z=1.5,\alpha_{xz}^y=0.2,\alpha_{xyz}^x=0.3,\alpha_{xyz}^y=0.15,\alpha_{xyz}^z=0.5$, especially for (c) $\alpha_y^z=-3.6,\alpha_{yz}^z=1.2$, (d) $\alpha_y^z=2.4,\alpha_{yz}^z=-1.85$, (e) $\alpha_y^z=1.1,\alpha_{yz}^z=-0.5$.
\\
\\

\section{The quantized Zak phase in spin-$\frac{1}{2}$ 1D Hamilton with mirror symmetry} 
\label{app:zak} 

In this section, we prove in a more rigorous method that the Zak phase of the 1-d Hamiltonian $H_{k_x,k_y}(k_z)$ with $\frac{1}{2}$-spin is quantized when the system is invariant under the mirror $m_z$ (and of course fully gapped). For brevity, we will omit the subscript $k_x,k_y$ and write $k_z$ as $k$ in the proving process. The $m_z$ symmetry requires that 
	\begin{equation}
		\phi_{n}(-k)=e^{i\theta_{nk}}\hat{m}\phi_{n}(k),
		\label{eq:phi_k}
	\end{equation}
	where $\phi_{n}(k)$ is the eigenstate of band $n$ and $\theta_{nk}$ is a continuous and periodic gauge. Equivalently, we have 
	\begin{equation}
		\phi_{n}(k)=e^{i\theta_{n,-k}}\hat{m}\phi_{n}(-k).
		\label{eq:phi_mk}
	\end{equation}
	By plugging Eq.~\ref{eq:phi_mk} into Eq.~\ref{eq:phi_k} and also considering $m_z^2=-1$ for spin-$\frac{1}{2}$ systems, we can get 
	\begin{equation}
		e^{i(\theta_{nk}+\theta_{n,-k})}=-1,
		\label{eq:theta_k_mk}
	\end{equation}
	$i.e.$, $\theta_{nk}+\theta_{n,-k}=(2u_{n}+1)\pi$ with integer $u_n$.

	The constraint imposed on the Berry connection $A_{nk}=i\langle\phi_{nk}|\partial_{k}|\phi_{nk}\rangle$ under the mirror symmetry can be formulated as 
	\begin{align}
		A_{n,-k} & =i\langle\phi_{n,-k}|\partial_{k}|\phi_{n,-k}\rangle\nonumber \\
	 		& =i\langle\phi_{n,k}|e^{-i\theta_{nk}}\hat{m}^{-1}\partial_{k}\hat{m}e^{i\theta_{nk}}|\phi_{n,-k}\rangle\nonumber \\
	 		& =-i\langle\phi_{n,k}|e^{-i\theta_{nk}}\partial_{k}e^{i\theta_{nk}}|\phi_{n,-k}\rangle\nonumber \\
	 		& =-A_{nk}+\partial_{k}\theta_{nk}.
	\end{align}
	Thus the Zak phase is
	\begin{equation}
		\mathcal{Z}_{n}=\int_{-\pi}^{\pi}A_{nk}\mathrm{d}k=\int_{0}^{\pi}\partial_{k}\theta_{nk}\mathrm{d}k=\theta_{n}(\pi)-\theta_{n}(0).
	\end{equation}

	Further the periodicity of Bloch states requires that $\theta_{n}(\pi)=\theta_{n}(-\pi)+2v_{n}\pi$ with integer $v_n$, together with the constraints from Eq.~\ref{eq:theta_k_mk} as
	\begin{align}
		2\theta_{n}(0) & =(2u_{n}+1)\pi\\
		\theta_{n}(\pi)+\theta_{n}(-\pi) & =(2u_{n}+1)\pi,
	\end{align}
	and we can conclude that the Zak phase of the 1-D spin-$\frac{1}{2}$ system with mirror symmetry is quantized $\mathcal{Z}_{n}=\theta_{n}(\pi)-\theta_{n}(0)=v_{n}\pi$.

\bibliographystyle{apsrev4-1}
\bibliography{Draft}

\end{document}